\documentclass[]{osa-article}

\journal{oe}

\usepackage[separate-uncertainty=false]{siunitx}
\DeclareSIUnit{\gauss}{G}

\usepackage{xspace}

\newcommand{\x}{$x$\xspace}
\newcommand{\y}{$y$\xspace}
\newcommand{\z}{$z$\xspace}

\newcommand{\od}{OD\xspace}

\newcommand{\na}{$^{23}$Na\xspace}
\newcommand{\ket}[1]{\vert#1\rangle}

\usepackage{xstring}
\newcommand*{\aref}[1]{%
	\IfBeginWith{#1}{eq:}{Eq.~\ref{#1}}{}%
	\IfBeginWith{#1}{fig:}{Fig.~\ref{#1}}{}%
	\IfBeginWith{#1}{tab:}{Table~\ref{#1}}{}%
	\IfBeginWith{#1}{appendix:}{Appendix~\ref{#1}}{}%
	\IfBeginWith{#1}{sec:}{Section~\ref{#1}}{}%
}

\graphicspath{{figures/}}

\usepackage{xcolor}

\articletype{Research Article}

\begin{document}

\title{Single-shot reconstruction of the density profile of a dense atomic gas}

\author{C. Mordini\authormark{1}, D. Trypogeorgos\authormark{1, 2, *}, L. Wolswijk\authormark{1}, G. Lamporesi\authormark{1, 2}, and G. Ferrari\authormark{1, 2}}

\address{\authormark{1} INO-CNR BEC Center and Dipartimento di Fisica, Universit\`a di Trento, 38123 Povo, Italy\\
\authormark{2} Trento Institute for Fundamental Physics and Applications, INFN, 38123 Povo, Italy}

\email{\authormark{*}d.trypogeorgos@unitn.it} 
\homepage{http://bec.science.unitn.it/} 


\begin{abstract}
Partial transfer absorption imaging (PTAI) of ultracold atoms allows for repeated and minimally-destructive measurements of an atomic ensemble.
Here, we present a reconstruction technique based on PTAI that can be used to piece together the non-uniform spatial profile of high-density atomic samples using multiple measurements.
We achieved a thirty-fold increase of the effective dynamic range of our imaging, and were able to image otherwise saturated samples with unprecedented accuracy of both low- and high-density features.
\end{abstract}


\section{Introduction}

Resonant absorption imaging (RAI), a broadly employed method for imaging cold atomic samples, measures the transmission of the atomic medium, which scales exponentially with the optical density (\od).
Atomic ensembles near/past the Bose-Einstein condensation (BEC) threshold can have extremely high peak densities $n_0\ge\SI{e20}{m^{-3}}$, which translates to typical peak ODs $\gtrsim 100$, way past the saturation limit of RAI~\cite{Ketterle1996making}.
Although for large BECs this problem persists even when imaging an expanded and dilute cloud, it is exacerbated for direct in-situ imaging~\cite{Ketterle1996making}.
Even in regimes of saturated absorption and high-intensity imaging~\cite{Reinaudi2007,HueckOptExpress.25.008670}, the dynamic range of RAI allows access to a range of ODs of no more than one order of magnitude.
Moreover, the process is fully destructive, since it imparts large kinetic energy to the BEC, and it allows for a single image per experiment; its use for measuring dynamic processes is limited by the cycle-time of the experiment, which is typically a few tens of seconds, and is only applicable in case of fully reproducible events.

Alternative imaging methods, such as phase-contrast~\cite{Andrews1997, Bradley1997}, Faraday imaging~\cite{Gajdacz2013,Higbie.95.050401}, and diffraction contrast imaging~\cite{Turner.72.031403}, take advantage of the high index of refraction of the BEC and give a dispersive signal.
Although these techniques are non-destructive, their dynamic range is still limited in a manner similar to RAI.

Here we demonstrate a new technique, based on partial transfer absorption imaging (PTAI)~\cite{Freilich2010}, that is capable of minimally-destructive imaging and has an exceedingly larger dynamic range.
PTAI is a versatile method that was used for in-situ imaging of superfluid flow in annular geometries~\cite{Ramanathan2011superflow,Ramanathan2012}, to observe the real-time dynamics of vortices~\cite{Serafini2015,Serafini2017,serafiniDynamicsVorticesTheir2017},
in-trap oscillations of a quantum gas~\cite{Seroka2019}, and measurement of the thermodynamic equation of state of an atomic gas~\cite{Mordini2020}.
It is implemented by coherently transferring a fraction of the atoms to an auxiliary energy level where it can be imaged by means of an optical cycling transition to an electronically excited state (\aref{fig:experiment}b).
The non-transferred atoms remain in the ground level which is far off-resonance from the optical transition and are left largely undisturbed by the imaging light.

Our technique accurately measures the in-situ density profile of condensed samples by taking several partial-transfer pictures of the same atomic sample, tuning the fraction of the outcoupled atoms so as to image areas of the BEC in different ranges of densities, always with optimum signal-to-noise ratio (SNR).
A reconstruction algorithm we developed, similar to high-dynamic-range (HDR) photography~\cite{mann1993compositing}, pieces together the information from the different pictures, allowing us to obtain a complete density profile in a single experimental realization of the condensate (shot).

\section{Experiment}
\begin{figure}
  \centering
    \includegraphics[]{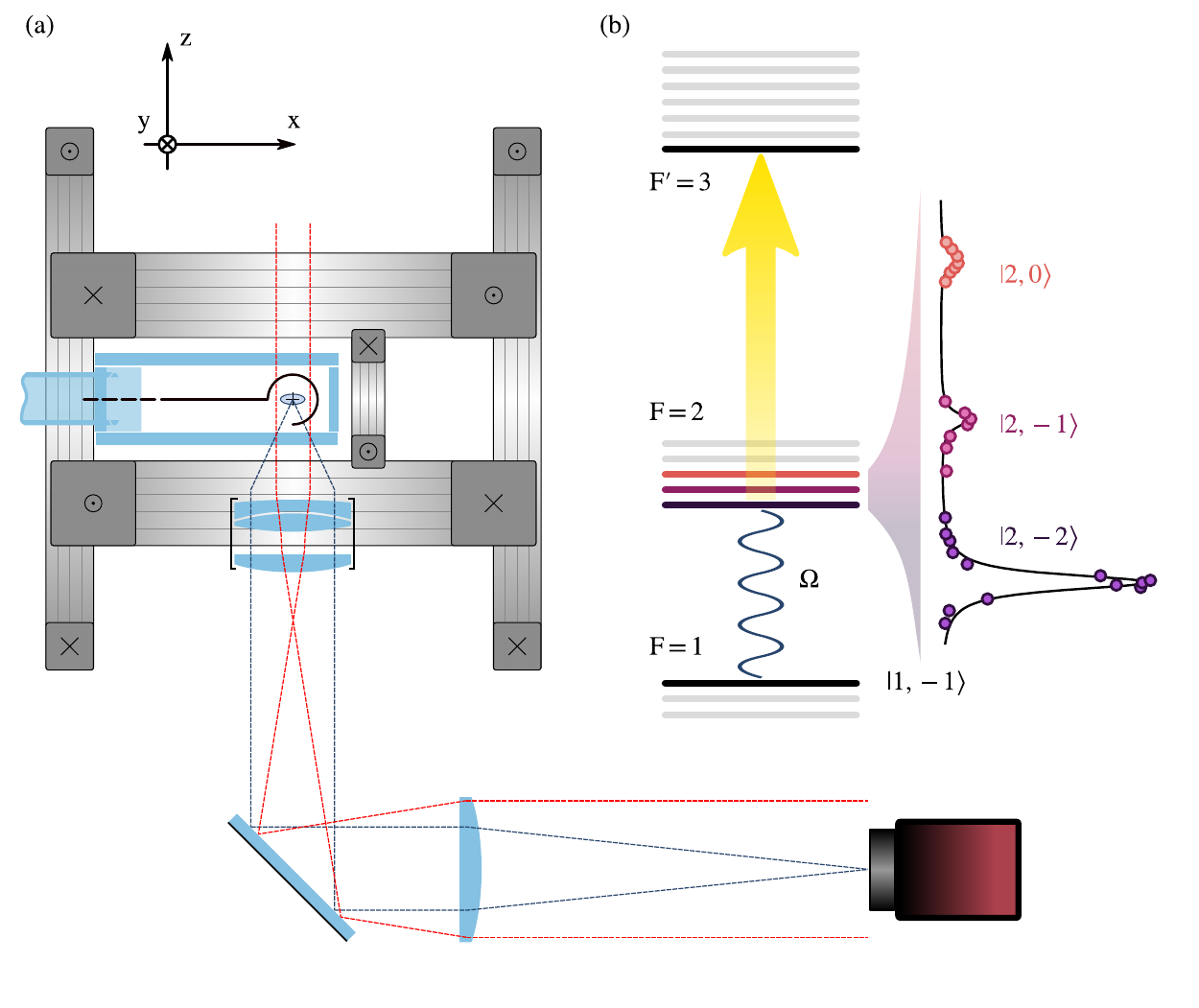}
  \caption{Experimental apparatus and atomic transitions. (a) The BEC (blue) is confined by the magnetic field generated by the five coils (gray). 
  It is imaged in absorption along the \z-axis using a relay optical system.
  The microwave field is generated by a hook antenna placed on the side of the glass cell.
  (b) The atoms, initially in $\ket{1,-1}$, are transferred via microwave radiation of Rabi frequency $\Omega$ to $\ket{2,-2}$ and are subsequently imaged in $F^\prime=3$.
  The spectroscopy (vertical) shows the three allowed microwave transitions between $\ket{1,-1}$ and the $F=2$ manifold. The separation between the peaks corresponds to the Larmor frequency of \SI{700}{\kilo\hertz}.
  }
  \label{fig:experiment}
\end{figure}

In our experiment we produce \na BECs with about $N=5\times 10^6$ atoms.
Our apparatus is described elsewhere~\cite{Lamporesi2012source}; here we focus only on the imaging part along the \z axis.
The sample is trapped in a magnetic Ioffe-Pritchard trap with an elongated geometry with trapping frequencies $\omega_x/2\pi = \SI{8.83 \pm 0.02}{\hertz}$, $\omega_y/2\pi = \omega_z/2\pi = \SI{100.8 \pm 0.7}{\hertz}$, and is polarized in the $\ket{F, m_F}=\ket{1,-1}$ hyperfine ground level.
A $|B|=\SI{1}{\gauss}$ bias magnetic field is applied to the system along \x, corresponding to a Larmor frequency of \SI{700}{\kilo\hertz}.
We implement PTAI using a $\approx \SI{1.77}{\giga\hertz}$ microwave field to transfer a fraction of the atoms to the upper hyperfine manifold $F=2$, that we subsequently image with light resonant to the $F = 2 \to F^\prime = 3$ cycling transition.
The microwave field is generated by a signal generator (Marconi Instruments 2024), amplified with a \SI{100}{\watt} amplifier (Minicircuits ZHL-100W-272+), and delivered to the atoms via a hook antenna (\aref{fig:experiment}a).

The microwave-field coupling induces Rabi oscillations between the two levels and the fraction of atoms transferred after a pulse time $t$ is
\begin{equation}
  \label{eq:rabi-fraction}
  P(t, \delta) = \left(\frac{\Omega}{\tilde\Omega(\delta)}\right)^2\sin^2\left(\frac{\tilde\Omega(\delta)\, t}{2}\right),
\end{equation}
with $\Omega$ the Rabi frequency and $\delta$ the detuning from resonance that lead to the system oscillating at the generalized Rabi frequency $\tilde\Omega = \sqrt{\Omega^2 + \delta^2}$.

We use $\ket{2,-2}$ as auxiliary state for the imaging, since this is the transition that has the largest Clebsch--Gordan coefficient, and spin-flipping collisions in the resulting hyperfine mixture are suppressed due to conservation of total angular momentum~\cite{Gorlitz2003}.
The choice of the auxiliary state is not crucial for the subsequent imaging, as the separation of the $F=2$ magnetic sublevels is of the order of the Larmor precession frequency and much smaller than the optical linewidth of \SI{10}{\mega\hertz}.
Figure~\ref{fig:experiment}b shows the population of the individual magnetic sublevels in the $F=2$ manifold with respect to the microwave frequency.

The magnetic trapping has an effect on the transfer, as it induces a spatially dependent detuning that we compute from the trap geometry. The equilibrium position of the BEC does not coincide with the minimum of the magnetic field, but is shifted downwards due to the gravitational sag, $z_{sag} = g/\omega_z^2$, where $g$ is the local acceleration of gravity.
This leads to the detuning
\begin{equation}
  \label{eq:uw-detuning-in-space}
  \hbar\delta(x, y, z) = \frac{3}{2}m \left( \omega_x^2 x^2 + \omega_y^2 y^2 + \omega_z^2 (z^2 - 2 z z_{sag}) \right)
\end{equation}
varying quadratically along all directions, where $m$ is the mass of \na and the factor 3 corresponds to the Bohr magneton difference between the two coupled states. The reference frame is centered on the atoms, where the microwave field is set on resonance and we have $\delta = 0$.

After the microwave extraction, we image the atoms in-situ using a relay imaging system with a magnification of 8.0(1) and a resolution of \SI{2}{\micro\meter}.
The probe light along the \z axis is absorbed resonantly by the atoms in the upper hyperfine manifold and casts a shadow in the camera (Allied Vision Stingray F-201B).
Combining the pictures with and without the atoms present, we reconstruct the OD of the sample from the ratio of the image counts (see Appendix A in~\cite{Ketterle1996making}).

\section{HDR reconstruction method}

\begin{figure}
  \centering
  \includegraphics[]{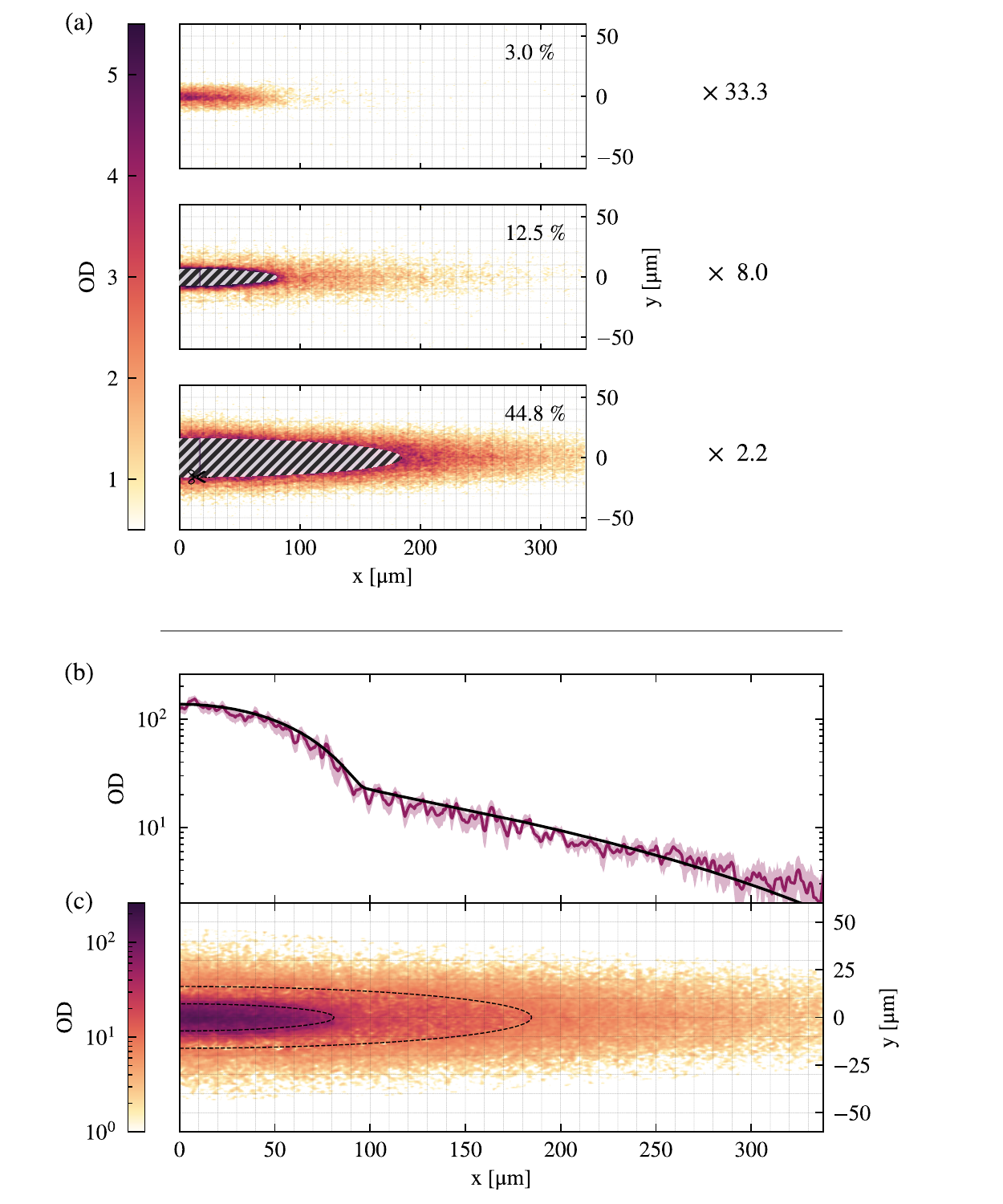}
  \caption{Reconstruction method.
  (a) Three partial-transfer images of the same BEC are taken in a single shot, and rescaled by the inverse of the extraction $P_{xy}$ in the imaging plane. The nominal scaling factors on the right correspond to $1/P_0$.
  The grid size is roughly equal to five times our resolution.
  (b, c) The resulting HDR image is the weighted mean of the above frames. The dotted ellipses correspond to the cropped-out regions above.
  The profile of a one-dimensional slice at the center of the density profile agrees well with the Hartree--Fock theory (black line).
  Notice that the OD spans two orders of magnitude in the reconstructed image.}
  \label{fig:hdr}
\end{figure}

For these measurements we produce condensates at a temperature of about \SI{230}{\nano \kelvin}, corresponding to a BEC fraction of about \SI{50}{\percent}.

We take a series of pictures of the same atomic cloud with increasingly longer microwave pulses.
In each picture, we choose the extracted fraction in order to bring a different part of the cloud to a level of OD optimal for our imaging parameters, that are appropriate for dense \na BECs~\cite{Horikoshi2017}.
Short pulses transfer a small fraction of atoms, whose spatial distribution is the same one as the original BEC and in the center has a peak OD $\approx 4$ which  can be imaged reliably with high-intensity RAI. The OD in the thermal tails, however, is still too low to be measured with a sufficient SNR.
Longer microwave pulses increase the apparent OD of the thermal part while the center of the cloud becomes too optically thick saturating the imaging (\aref{fig:hdr}).
The inhomogeneous field of the magnetic trap introduces a spatial dependency in the local extraction $P$, since $\delta$ depends on space.
The effect of the detuning, integrated along the imaging direction, is minimized with a large value of $\Omega$.
We work at $\Omega/2\pi = \SI{59.2 \pm 0.3}{\kilo\hertz}$, that gives a relative error in the OD $< \SI{5}{\percent}$ (see \aref{sec:non-uniform}).
For given $\Omega$ and $t$, we define a nominal extraction $P_0 = P(t, \delta = 0)$ as the fraction extracted at resonance, which is realized in the center of the atomic cloud.

Using a method inspired by HDR photography, we combine the information from different frames to reconstruct a complete image of the optical density of the trapped sample.
Each frame is fitted with a bimodal distribution and re-centered, so that the origin of the coordinates coincides with the common center of all the imaged atomic clouds.
We crop the regions in the pictures that are above the saturation threshold OD = 4, except for the smallest extraction where the peak OD is already below the threshold.
This value, dictated by the imaging conditions, is the one at which the transmitted probe light becomes comparable with the camera noise.
The cropping mask is the convex hull of the biggest simply-connected region formed by the pixels above threshold.
This choice ensures that the edges of the cropped region are smooth and avoids that the crop is biased by noise in the OD around the threshold value.
We account for the spatial variation of $P$ in the imaging plane by rescaling each picture by $P_{xy} = P(t, \delta(x, y, z=0))$, the local extraction in the $xy$ plane calculated from Eqs.~\ref{eq:rabi-fraction} and \ref{eq:uw-detuning-in-space}, and overlay the frames on top of each other.
The good match of the rescaled OD in the overlap region is evidence of accurate calibration of the Rabi frequency.
Finally, we average the different frames weighting them by their SNR, where the signal is evaluated from the bimodal distribution fitted before rescaling, and the noise level is the same in all the frames. This favors the frames with higher extraction and maximizes the SNR of the reconstructed image.

Figure~\ref{fig:hdr} shows a pictorial representation of the method.
In the top panel there is a stack of three different frames, imaged with increasing microwave pulsetimes (from top to bottom) \SI{0.9}{\us}, \SI{1.94}{\us}, \SI{4}{\us}, which correspond to a nominal extraction of \SI{3}{\percent}, \SI{12.5}{\percent}, \SI{44.8}{\percent} respectively.
The complete reconstruction of the atomic density is shown in \aref{fig:hdr}c and corresponds to the weighted average of the rescaled pictures above.
Along the long axis of the trap, we fit the OD to a Hartree--Fock profile~\cite{Dalfovo1999} to verify that our reconstruction algorithm leads to physically meaningful results (see \aref{fig:hdr}b). The shaded region represents a 1-$\sigma$ errorbar including both the statistical noise coming from the imaging process, related to shot noise in the camera counts, and the systematic error due to the uncertainty in the Rabi frequency and the effect of the field inhomogeneity along \z. All those error sources are uncorrelated and are added in quadrature.

\section{Calibration of the Rabi frequency}
\label{sec:calibration}

The implementation of our HDR method requires the knowledge of $\Omega$ and $\delta$, the two parameters appearing in \aref{eq:rabi-fraction}, to precisely determine the in-situ OD of the atomic sample.
As $\delta$ is known from the trap geometry, the only remaining parameter is the Rabi frequency $\Omega$ which we obtain in a single shot using anew PTAI.

The Rabi frequency can be measured by simply following the Rabi flopping of the system (see \aref{fig:rabi_flop}a).
This is normally done by pulsing the coupling field for a given time $t$ and then measuring both the ground- and excited state populations using a Stern--Gerlach technique to minimize errors in the Rabi frequency due to shot-to-shot number fluctuations in the preparation of the atomic sample.
In our case, we implement this by letting the transferred atoms fall under the action of gravity and the antitrapping magnetic field.
We image the falling atoms after a time-of-flight (TOF) of $\sim\SI{10}{\milli\second}$ and the remaining atoms are then released from the trap and imaged in TOF with an optical repumper.

\begin{figure}
  \centering
  \includegraphics[]{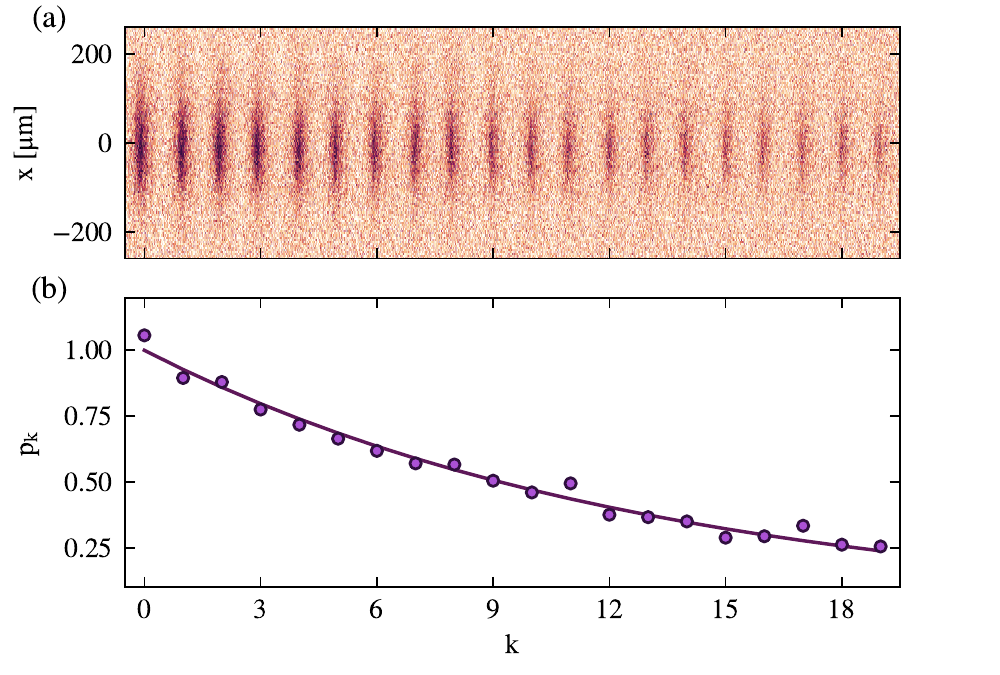}
  \caption{Single-shot measurement of the Rabi frequency. (a) A composite image showing a sequence of extractions from a single realization of the atomic sample. A constant fraction of atoms is extracted every \SI{16}{\milli\second}.
  (b) The relative atom number progression is modeled with a geometric series (solid line) from which we infer the value of the Rabi frequency.}
  \label{fig:rabi_geometric}
\end{figure}

However, the dynamics of the system is actually governed by the generalized Rabi frequency $\tilde\Omega$.
A non-uniform detuning (such as the one present in a magnetically trapped sample) leads to a locally varying precession rate which in practice reduces the spatial coherence of the oscillations (see \aref{sec:non-uniform}).
Maximising $\Omega$ allows us to partially circumvent this limitation by reducing the duration the coupling field is applied for.
From \aref{eq:rabi-fraction} it follows that for $\tilde\Omega t \ll 1$ the transferred fraction $\Omega^2 t^2 / 4$ is independent of the microwave frequency.
For a given product $\tilde\Omega t$, a large Rabi frequency reduces the distortions, as the detuning term $\delta \ll \Omega$ becomes negligible due to power-broadening of the resonance and the transfer fraction is now proportional to $\Omega$.

To this end we devised an alternative method to measure $\Omega$ which can also be done in a single-shot, dramatically increasing the measurement rate.
We apply a sequence of microwave pulses coupling to $\ket{2, -2}$ to extract a small fraction of the trapped sample every \SI{16}{\ms}.
We repeatedly image the extracted atoms after a \SI{4}{\ms} time-of-flight (TOF) using a camera along \y (not shown in \aref{fig:experiment}), for up to 20 times before the BEC is depleted, recording the atom number over time (see \aref{fig:rabi_geometric}a).
Applying always the same microwave pulse, the number of atoms in the cloud decreases by a constant fraction $q = \sin^2(\Omega t /2)$ at each extraction.
With $N_k$ the number of atoms extracted after $k+1$ pulses, the relative number of extracted atoms $p_k = N_k / N_0$ follows a geometric series $p_k = (1 - q)^k$.
From the series shown in \aref{fig:rabi_geometric}b we obtain $q$, which leads to a Rabi frequency of \SI{58.8 \pm 0.5}{\kilo \hertz} for a $t=\SI{1.5}{\micro\second}$ pulse, that is consistent with the value measured by following the Rabi flopping in multiple shots, with the advantage of being far less sensitive to the frequency detuning between the coupling field and the individual resonance of each atom.
In fact this method allows to considerably reduce the duration of the pulses for the same intensity of the applied microwave field, hence suppressing the effects of the local detuning due to the Fourier spectral broadening of the applied pulse.

\section{Considerations}

\subsection{Non-uniform magnetic field}
\label{sec:non-uniform}

\begin{figure}
  \centering
  \includegraphics[]{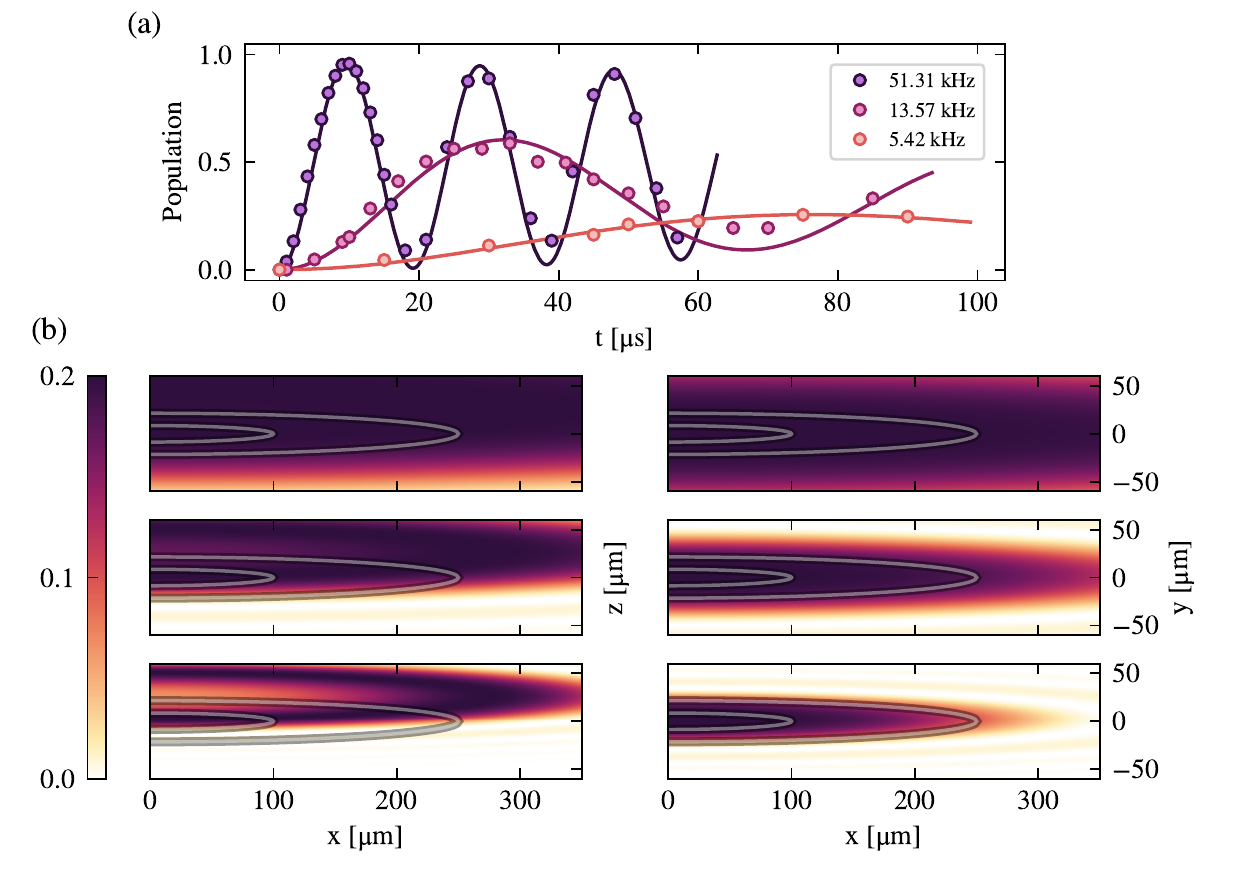}
  \caption{Non-uniform Rabi flops: effect of the Rabi frequency. (a) The population of the upper state oscillates at less than full amplitude depending on the value of $\Omega$ due to the spatial dependence of $\delta$.
  (b) Spatial profile of the transferred population on the $xz$ plane (left) and on the $xy$ plane (right) for the different values of the Rabi frequency shown in (a), \SI{51.31}{\kilo\hertz}, \SI{13.57}{\kilo\hertz}, \SI{5.42}{\kilo\hertz} from top to bottom. 
  The gray ellipses show the extent of the BEC (inner) and the thermal cloud (outer).}
  \label{fig:rabi_flop}
\end{figure}

As mentioned above, a high Rabi frequency and short pulse times (i.e., small extractions) both contribute to suppress the effect of $\delta$ in \aref{eq:rabi-fraction} and obtain an extraction as close as possible to $P_0$ at every point in space.

Figure~\ref{fig:rabi_flop} shows the effect of the choice of $\Omega$. In (a) we show Rabi flops of the atomic population measured with the traditional Stern--Gerlach method for different microwave power.
The reduced contrast in the Rabi flop is an effect of the non-uniform detuning.

The values of the Rabi frequency are extracted by fitting the transferred population with
\begin{equation}
  \label{eq:rabi-fraction-fit}
  \tilde P(t) = \frac{1}{2\sqrt{1 + 2D^2}} \left( 1 - \frac{\cos(\tau + \arctan(b) / 2)}{(1 + b^2)^{1/4}} \right),
\end{equation}
where $\tau = \Omega\, t$, $D = \Delta_0/\Omega$ and $b = \tau D^2 / (1 + 2D^2)$. $\Delta_0$ is an effective range of detuning spanned by the condensate, and depends on the microwave field strength and the size of the atomic sample.
This is an approximate model which captures the effect of the vertical field gradient (see Appendix A). 
We observe how higher values of $\Omega$ reduce the effect of the inhomogeneous term and increase the contrast of the oscillations.
In \aref{fig:rabi_flop}b we compute the spatial profile of the transferred fraction in the $xz$ plane, for the different values of $\Omega$ shown in (a) and a nominal extraction $P_0 = 0.2$.
The atomic sample occupies an area delimited by the gray ellipses, with semi-axes equal to the Thomas--Fermi radii of the BEC (inner) and $2\,\sigma$, where $\sigma$ is the width of the thermal cloud (outer).
Here we see that for $\Omega / 2 \pi > \SI{20}{\kilo\hertz}$, which is close to the value of $\delta$ at the edges of the BEC, we achieve a nearly uniform extraction in the region occupied by the atoms. For the three reported values of the Rabi frequency in \aref{fig:rabi_flop}, the relative variations in the extraction between the center and the lower side of the thermal wings, $(P(0,0,z = -2\,\sigma) - P_0) / P_0$, are  \SI{-7.6}{\percent}, \SI{-72}{\percent} and \SI{-95}{\percent}, from top to bottom.

\begin{figure}
  \centering
  \includegraphics[]{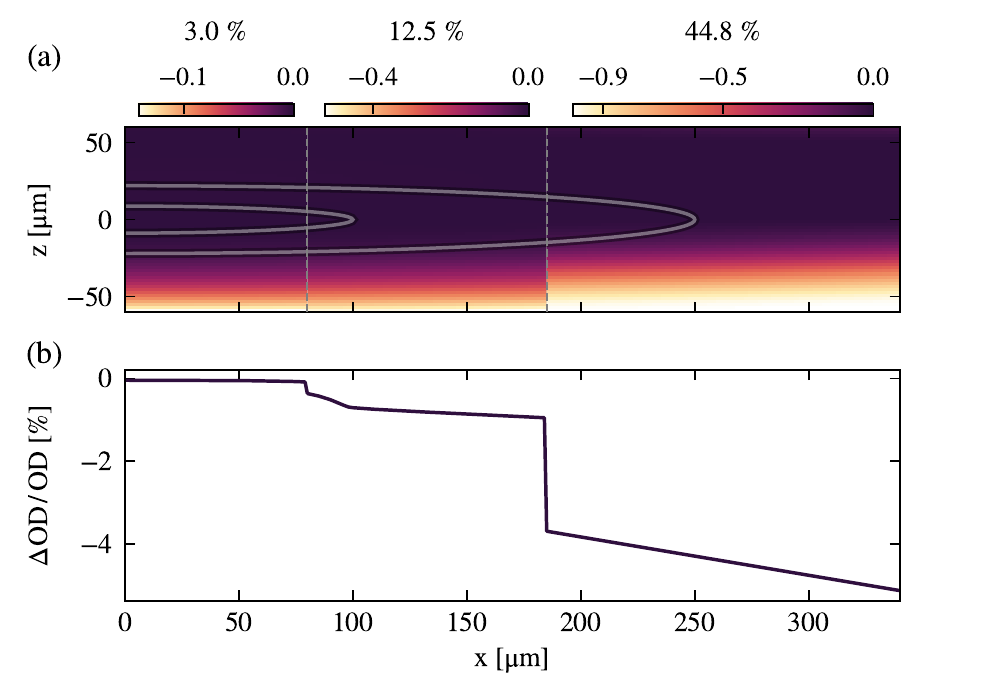}
  \caption{Non-uniform Rabi flops: effect of the pulse time.
  (a) Map of $(P(t, \delta) - P_{xy}) / P_{xy}$ in the $xz$ plane, computed for the parameters used in the reconstruction. The three sections correspond to the non-cropped regions of the three frames. In each one the colormap spans a different range, to highlight the very different extents in the spatial variation of the extraction.
  (b) Relative error in the OD along the \x axis due to the field inhomogeneity.
  The gray ellipses show the extent of the BEC (inner) and the thermal cloud (outer).}
  \label{fig:inhomog}
\end{figure}

Figure~\ref{fig:inhomog} shows the effect of the microwave pulsetime, and quantifies the systematic error introduced by the magnetic field in the reconstruction procedure.
In \aref{fig:inhomog}a we plot a map of $(P(t, \delta) - P_{xy}) / P_{xy}$, the relative difference between the actual value of the extraction and the one used in our rescaling, along the $xz$ plane. This is calculated for a Rabi frequency of \SI{59.2}{\kilo\hertz} and different pulsetimes, the same parameters used in the reconstruction of \aref{fig:hdr}.
The three sections correspond to the regions where each frame in \aref{fig:hdr}a, identified by the respective values of $P_0$ quoted above, contributes to the reconstructed image.
We see that for the smallest extraction, which is relevant for the central part of the atomic sample, the field-induced detuning changes the extraction by almost \SI{10}{\percent} only in regions far from the dense BEC. For the highest extraction, instead, the effect of the detuning is much stronger, but this is compensated by the extremely reduced value of the atomic density.

This leads to a systematic error in the measured OD
\begin{equation}
    \Delta OD = \int \frac{P(t, \delta) - P_{xy}}{P_{xy}}\ n\, dz
    \label{eq:delta_od}
\end{equation}
that we evaluate integrating the extraction profiles along \z, and using as density profile the Hartree--Fock fit of the reconstructed OD in \aref{fig:hdr}c.
Figure~\ref{fig:inhomog}b shows the ratio $\Delta OD / OD$, that is the relative error on the optical density along the \x axis.
The contribution to the integral in \aref{eq:delta_od} from the regions far from the $xy$ plane, where the error in the extraction is stronger, is highly suppressed by the reduced value of the density. As a result, the error in the OD remains below \SI{1}{\percent} in the whole region of the condensate, and $< \SI{5}{\percent}$ in the thermal wings.
This result holds also along the \y axis after proper rescaling of the coordinates, given the elliptical shape of the trapping potential. Therefore, we have that the relative error is $< \SI{5}{\percent}$ on the whole reconstructed image.

\subsection{Collective mode excitation}

The excitation of collective modes can potentially distort the reconstructed density profile of the BEC.
The dipole mode, simply translates the BEC without distorting it; the BEC can be re-centered as long as the translation is perpendicular to the imaging axis but might otherwise introduce focusing errors.
Higher order modes, e.g. the quadrupole mode, are naturally excited during the extraction process~\cite{Serafini2015} and distort the shape of the BEC at timescales close to the inverted trapping frequencies.
It is possible to work around this effect by arranging for the sampling frequency to be equal to the mode frequency so all subsequent images are consistent with each other.

\begin{figure}
  \centering
  \includegraphics{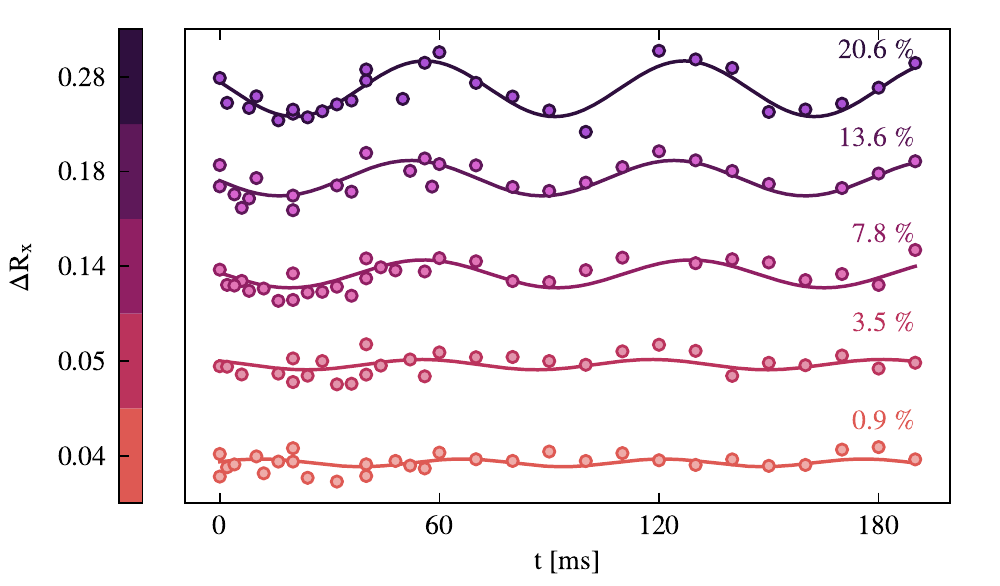}
  \caption{Quadrupole mode excitation. The Thomas--Fermi radius oscillates in time after a single extraction has taken place.
  The oscillations for different extraction fractions are shifted vertically for clarity.
  The curves are colored based on the relative value of their oscillation amplitude $\Delta R_x$, which is directly proportional to the extraction fraction.
  The quadrupole mode becomes appreciable for extraction fractions larger than \SI{5}{\percent}.}
  \label{fig:excitation}
\end{figure}

However, avoiding its excitation altogether is preferable.
Moreover, if the sampling frequency is much faster than the mode frequency the shape of the BEC can be considered essentially frozen.
Figure~\ref{fig:excitation} shows the variations of the Thomas--Fermi radius along the \x axis, $R_x$, normalized to its value before any extraction takes place, for different extraction fractions.
The relative amplitude of the excited quadrupole mode $\Delta R_x$ is smaller than 0.05 for extractions smaller than \SI{3.5}{\percent}.
Motivated by this observation, we arranged the sequence of images in \aref{fig:hdr} as follows.
The \SI{3.5}{\percent} image is taken at $t=0$ and leaves the BEC largely undisturbed.
Then the \SI{12.5}{\percent} and \SI{44.8}{\percent} images are taken 70\,ms later but only \SI{600}{\micro\second} apart, which is much smaller than the period of the quadrupole mode.
Within these \SI{600}{\micro\second} we shine a \SI{5}{\micro\second} resonant pulse along \y that pushes the already imaged atoms along a transverse direction thus removing them from the field of view.

\subsection{Off-resonant scattering}

Acquiring multiple measurements normally requires an equal number of atom-light interactions with the imaging laser resonant to the $F=2 \to F^\prime = 3$ transition.
An imaging laser pulse follows the transfer of atoms from $F=1 \to F=2$.
The laser light interacts off-resonantly with atoms in the otherwise shelved $F=1$ state, which for \na is \SI{1.77}{\giga\hertz} detuned, and leads to the reservoir of atoms being depleted~\cite{grimmOpticalDipoleTraps2000}.
This introduces a systematic error in the measurement of the atom number for the subsequent images which can be measured and accounted for.
For our typical imaging intensity $I/I_{\textrm{sat}} = 4$, where $I_{\textrm{sat}} = \SI{13.4}{mW/cm^2}$ is the saturation intensity of \na, the off-resonant scattering rate is \SI{e3}{s^{-1}}, which leads to negligible losses for a \SI{5}{\micro\second} imaging pulse.

Incidentally, the non resonant scattering from $F=1$ atoms also affects the apparent number count of $F=2$ atoms for slightly out-of-focus systems~\cite{Wigley2016}.

\section{Conclusions}

We developed a powerful single-shot, HDR atomic imaging method which allows to increase the dynamic range of the imaging by more than one order of magnitude and accurately reconstruct the true OD of the atomic sample.
The fraction imaged within each PTAI snapshot is optimized to different ranges of the OD, while the merging of the information from different snapshots is implemented thanks to the accurate knowledge of the extracted fraction.
The final outcome is the complete optical density distribution.
We also devised a procedure for measuring the Rabi frequency of the microwave transitions which is highly insensitive to non-uniformities across the atomic system.
The method is based on the iterated application of short microwave pulses of constant duration and is benchmarked against usual Rabi flopping.

We expect our HDR reconstruction method to be widely applicable to a number of similarly dense atomic systems like, for instance, quantum droplets~\cite{Chomaz2016,cabreraQuantumLiquidDroplets2018,Semeghini2018}.
We have already successfully used it to measure the equation of state of a three-dimensional bosonic gas~\cite{Mordini2020}.

\appendix

\section*{Appendix A: Non-uniform Rabi model}
\label{appendix:A}

After the application of a microwave pulse of duration $t$, the transferred fraction to $F=2$ is
\begin{equation}
  \label{eq:p-tilde}
  \tilde P(t) = \frac{N_2}{N} = \frac{1}{N} \int P(t, \delta(r))\ n(r)\, d^3r,
\end{equation}
where $N=N_1+N_2$ is the total number of atoms.
The probability distribution of \aref{eq:rabi-fraction} is weighted by the atomic density and spatially integrated.

To extract the value of $\Omega$ from a Rabi flop, we introduce some approximations.
Since the leading term in the magnetic field profile, within the extent of the atomic cloud, is the linear gradient along \z, we neglect the smaller quadratic terms in 
\aref{eq:uw-detuning-in-space} and rewrite $\delta \propto 2 z z_{sag}$.
We approximate the atomic density distribution with a Gaussian shape and integrate out the directions \x and \y where the detuning remains constant.
Equation~\ref{eq:p-tilde} now becomes
\begin{equation}
  \label{eq:p-tilde-approx}
\tilde P(t) \simeq \frac{1}{\sqrt{2\pi}\Delta_0}\int P(t, \delta)\ e^{-\delta^2/2\Delta_0^2}\, d\delta,
\end{equation}
which explicitly shows that the effect of the field gradient is to average the transferred population over the local detuning.
In our trap, a BEC of \num{5e6} atoms has a transverse Thomas-Fermi radius of \SI{12}{\micro \meter}, which corresponds to an effective span of detuning $\Delta_0=\SI{20}{\kilo \hertz}$.

Within these approximations the integral in \aref{eq:p-tilde-approx} can be solved analytically, resulting in \aref{eq:rabi-fraction-fit}.

\section*{Acknowledgements}
We are grateful to A. Farolfi for the many valuable discussions, and for his insightful contributions to the reconstruction algorithm and the data analysis.
We thank the Q@TN initiative.

\section*{Funding}
This work was supported by the project NAQUAS of QuantERA ERA-NET Cofund in Quantum Technologies (Grant Agreement N. 731473) implemented within the EU Horizon 2020 Programme, and by the Provincia Autonoma di Trento.

\section*{Disclosure}
The authors declare no conflicts of interest.

\bibliography{ptai.bib}

\end{document}